\documentclass[aps,prd,twocolumn,showpacs,showkeys,amsmath,amssymb,preprintnumbers]{revtex4-1}

\usepackage{graphicx}
\usepackage{epsfig}
\usepackage{bm}
\usepackage{float}
\usepackage{amsfonts}
\usepackage{graphicx}
\usepackage[dvips]{hyperref}
\usepackage{bm}
\usepackage{amsmath}
\usepackage[all]{xy}
\def\beq{\begin{eqnarray}}

\def\eeq{\end{eqnarray}}

\begin{document}

\title[Bulk-Boundary interaction and the second law in Ho\v{r}%
ava-Lifshitz cosmology]{Bulk-Boundary interaction and the second law in Ho\v{r}%
ava-Lifshitz cosmology}

\author{Miguel Cruz}
\email{miguel.cruz@ucv.cl}
\address{Facultad de F\'\i sica, Universidad Veracruzana, 91000 Xalapa, Veracruz, M\'exico and 
Instituto de F\'\i sica, Facultad de Ciencias, Pontificia Universidad Cat\'olica de Valpara\'\i so, 
Av. Brasil 4950, Valpara\'\i so, Chile}

\author{Samuel Lepe}
\email{samuel.lepe@pucv.cl}
\address{Instituto de F\'\i sica, Facultad de Ciencias, Pontificia Universidad Cat\'olica de Valpara\'\i so, 
Av. Brasil 4950, Valpara\'\i so, Chile}

\author{Francisco Pe\~na}
\email{francisco.pena@ufrontera.cl}
\address{Departamento de Ciencias F\'\i sicas, Facultad de Ingenieria y Ciencias, Universidad de la Frontera, Casilla 54-D, Temuco, Chile}

\begin{abstract}
By defining $Q$ as a function which realizes the energy transference between
the bulk and the boundary of spacetime, as we interprete it here in the
framework of Ho\v{r}ava-Lifshitz (HL) cosmology, flat case, we discuss the validity of the 
second law of thermodynamics in the light of the sign changes of $Q$ (changes in
the direction of energy transference) and its consequences through the
cosmic evolution, in particular, whether the thermal equilibrium between bulk and boundary
is reached or not. Additionally, we discuss possible phase transitions experienced by the bulk 
and the boundary (seen as sign changes in their heat capacities) through the cosmic evolution. The 
energy density in the bulk is modeled under an holographic perspective. As far as we know, 
currently there is not observational data on bulk-boundary interaction. 
\end{abstract}



\maketitle

\date{\today}

\section{Introduction}
\label{sec:intro}

An open question is whether the Universe is or not a thermodynamical system.
If this were true, the second law must be satisfied by any cosmological model 
that attempt to coherently describe the evolution.
On the other hand, if the Universe is not a thermodynamical entity, it is 
still an interesting problem to explore. If the Universe admits a thermodynamical 
description, we must also speak of its physics on thermal equilibrium. But 
it seems very difficult to consider equilibrium and the possibility 
to attain it, for instance, between different present species (under interaction or not). 
As we will see, an analogue situation is verified in HL cosmology when we think about 
bulk-boundary interaction.

Bulk-boundary interaction can lead to a new perspective for studying the
cosmic evolution, and the HL cosmology \cite{Mukohyama1} provides a good scenario 
to develop it \cite{Lepe2}. In this work, we analize the thermodynamical aspects 
for the aforementioned interaction where the emphasis will be focused on the thermal
equilibrium, if exists, and the second law. A crucial point is to analize if the second 
law is satisfied or not in this cosmology; an holographic philosophy will be used for
modeling the energy density present at the bulk. $8\pi G=$ $c=1$ units will be used 
throughout this work.

The paper is organized as follows: in Section II we will discuss briefly
some aspects of the function $Q$. In Section III,  considering the framework of the 
HL cosmology (flat case) we discuss the bulk and boundary temperatures and the second
law. In Section IV, we study the model by introducing an holographic scheme for the 
energy density present at the bulk. In Section V, we discuss the
phase transitions experienced by the bulk and the boundary of the spacetime
through the cosmic evolution. Finally, Section VI is devoted to conclusions.

\section{$Q$-function}
\label{sec:Q}

When two interacting fluids are considered, different Ansatzes
for $Q$ (this function measures the energy transference
between both fluids) can be seen in the literature, for instance,
\begin{equation}
Q=3\lambda H\rho,
\label{eq:1} 
\end{equation}%
where $\lambda $ is a constant parameter to be determined by observations, and
this $Q$ does not experience any sign changes through the evolution \cite{He4}. Either
way, the ``Ansatz philosophy'' for the function $Q$, is a first approximation to describe
the interaction according to the observational data \cite{He4, Salvatelli5}. In the present
work, we will use Eq. (\ref{eq:1}) as an example of interaction function between the
bulk and boundary, and later the case of a $Q$-function which exhibits sign changes.
Another approach, used here, is based on the holographic philosophy \cite{Lepe2, Arevalo6, Lepe7},
where we can obtain an explicit form for $Q$ after using some
parametrization over $\omega$ (equation of state parameter) or $q$
(deceleration parameter). For this case, it is possible to visualize sign
changes in $Q$, a fact that is interesting if we are thinking, for instance,
about eventual phase transitions experienced by the bulk or the boundary of the
spacetime. In reference \cite{Lepe7} were found some phase transitions
(sign changes in the heat capacity) experienced by two interacting fluids,
dark energy under an holographic scheme and dark matter considered as a
presureless fluid.

\section{Bulk/Boundary interaction and thermodynamics}
\label{sec:bbt}

In the framework of the flat HL gravity we study its cosmology and the 
projectable version of this theory preserves the diffeomorphism 
invariance \cite{Mukohyama1}. Here, the dynamical equation governing the 
cosmology is given by
\begin{equation}
\eta \left( 2\dot{H}+3H^{2}\right) =-p=-\omega \rho,
\label{eq:2} 
\end{equation}%
and the energy density satisfies the non-conservation equation
\begin{equation}
\dot{\rho}+3H\left( 1+\omega \right) \rho =-Q,
\label{eq:3} 
\end{equation}%
where $\rho $ represents the energy density present at the bulk, $\eta $ is
a dimensionless constant parameter associated to diffeomorphism invariance
(this parameter is well confined to the range $0<\eta <1$, see Ref. \cite{Lepe2}) and $Q$ the
interaction term. In the HL cosmology, $Q$ comes from the theory as an
``integration constant'' and it is not imposed by hand as done for the 
treatment of two interacting fluids. In Ref. \cite{Lepe2} was shown that in 
the HL cosmology $Q$ does not vanish at late times; in fact, it never vanishes. Therefore, 
the thermal equilibrium between the bulk and the boundary of the
spacetime can not be reached. In consequence, we do not share the idea that equilibrium can 
take place during late times or the idea of a conservation equation for $\rho$ as done in
Ref. \cite{Jamil3}. An interesting discussion over restrictions on the thermal
equilibrium between dark energy and cosmological horizon can be seen in
Ref. \cite{Poitras8}.

$\bullet$ Bulk temperature. \\
The Gibbs equation for the fluid at the bulk reads
\begin{equation}
T_{b}dS_{b}=d\left( \rho V\right) +pdV=\rho V\left( \frac{dV}{V}+\frac{d\rho 
}{\rho }\right),
\label{eq:4}   
\end{equation}
and by using the deceleration parameter defined by $q=-\left( 1+\dot{H}
/H^{2}\right) $ and $V=\left( 4\pi /3\right) H^{-3}$ (being $H^{-1}$ the radius
of the Hubble horizon), we obtain
\begin{equation}
T_{b}\dot{S}_{b}=\rho V\left[ 3(1+\omega )\left( 1+q\right) H+\frac{\dot{\rho%
}}{\rho }\right],
\label{eq:5} 
\end{equation}
so that, after using Eqs. (\ref{eq:2}) and (\ref{eq:3}), we have
\begin{equation}
T_{b}\dot{S}_{b}=\frac{8\pi }{3}\eta \left( \frac{q-\frac{1}{2}}{\omega }\right) %
\left[ 3(1+\omega )q -\frac{3\omega }{2\left( q-\frac{1}{2}\right) } \left( \frac{Q}{%
3\eta H^{3}}\right) \right].
\label{eq:6} 
\end{equation}
By using the integrability condition 
\begin{equation*}
\frac{\partial ^{2}S_{b}}{\partial
T_{b}\partial V}=\frac{\partial ^{2}S_{b}}{\partial V\partial T_{b}}, 
\end{equation*}
we can obtain the evolution equation for the bulk temperature \cite{Maartens8}
\begin{eqnarray}
\frac{dT_{b}}{T_{b}}&=&-3H\left( \frac{\partial p}{\partial \rho }\right) dt=3%
\frac{dz}{1+z}\omega \rightarrow T_{b}\left( z\right) = \nonumber \\
&=& C\exp \left(3\int \frac{dz}{1+z}\omega _{eff}\right),
\label{eq:7} 
\end{eqnarray}
where $\omega _{eff}$ is given by
\begin{equation}
\omega _{eff}=\omega \left[ 1+\frac{1}{2\left( q-1/2\right) }\left( \frac{Q}{%
3\eta H^{3}}\right) \right],
\label{eq:8} 
\end{equation}
where we have used again the Eqs. (\ref{eq:2}) and (\ref{eq:3}). So, we have
\begin{equation}
T_{b}\left( z\right) = C_{0}\exp \left( 3\int \frac{dz}{1+z}\omega \left[ 1+ 
\frac{1}{2\left( q-1/2\right) }\left( \frac{Q}{3\eta H^{3}}\right) \right]
\right),
\label{eq:9} 
\end{equation}
where $C_{0}$ is a constant. In Section (\ref{sec:IV}) we will show the form of Eq. (\ref{eq:9}) 
by using different choices for $Q$.

$\bullet$ Boundary temperature.\\ 
According to the holographic principle, the Hubble horizon (the boundary) 
has a temperature given by the simple expression
\begin{equation}
T_{h}=\frac{H}{2\pi},
\label{eq:10} 
\end{equation}
and the associated entropy given by \cite{Rong9}
\begin{equation}
S_{h}=8\pi ^{2}H^{-2}\rightarrow T_{h}\dot{S}_{h}=8\pi \left( 1+q\right).
\label{eq:11} 
\end{equation}
In Section (\ref{sec:IV}), we show that by using in Eq. (\ref{eq:2}) an holographic energy density, we obtain 
the following expression for $T_{h}\left( z\right) $
\begin{align}
& \frac{T_{h}\left( z\right)}{T_{h}\left( 0\right)} = \left( 1+z\right) ^{3/2} \times \nonumber \\
&\times \exp
\left( q_{1}\left[ \frac{1}{1+z}\left( \frac{1+\bar{z}}{2\left( 1+z\right) }
-1\right) -\left( \frac{1+\bar{z}}{2}-1\right) \right] \right),
\label{eq:12}
\end{align}
where we have used the parametrization $q\left( z\right) =1/2+q_{1}\left( z-\bar{z}
\right) /\left( 1+z\right) ^{2}$ \cite{Gong10}, where both quantities $q_{1}$ and $\bar{z}$ are
defined positives. Then, we have $T_{b}\left( z\right)$ from the HL
cosmology given by Eq. (\ref{eq:9}) and $T_{h}\left( z\right) $ from an holographic scheme given by 
Eq. (\ref{eq:12}). Under this scope we will study the second law.

$\bullet$ Second law. \\
According to Eqs. (\ref{eq:6}) and (\ref{eq:11}), we can write
\begin{widetext}
\begin{equation}
\dot{S}_{b}+\dot{S}_{h}\geq 0\rightarrow \frac{8\pi }{T_{b}}\eta
\left( \frac{q-\frac{1}{2}}{\omega }\right) \left[ \left( 1+\omega \right) q-\frac{
\omega }{2\left( q-\frac{1}{2}\right) }\left( \frac{Q}{3\eta H^{3}}\right) \right] +
\frac{8\pi }{T_{h}}\left( 1+q\right) \geq 0,
\label{eq:13}
\end{equation}
\end{widetext}
where we have to consider the following two conditions: for $Q>0$ the energy flow goes from the bulk to
the boundary of spacetime and for $Q<0$ the energy flows in opposite direction.
Now, from Eq. (\ref{eq:13}) we write
\begin{equation}
\frac{Q}{3\eta H^{3}}\leq  2\left[ \left( \frac{q-\frac{1}{2}}{\omega }\right)
q\left( 1+\omega \right)+ \left( 1+q\right) \frac{1}{\eta }\left( \frac{T_{b}%
}{T_{h}}\right) \right],
\label{eq:14}
\end{equation}
and, by recalling Eq. (\ref{eq:2}), $\rho H^{-2}/2\eta =\left( q-\frac{1}{2}\right) /\omega >0$,
we must have $q>1/2$ and\ $\omega >0$ or $q<1/2$ and\ $\omega <0$ always. In
Section (\ref{sec:IV}) we will compare both members of Eq. (\ref{eq:14}), for different choices
of the function $Q$ in order to verify the second law (whether or not the
inequality given in Eq. (\ref{eq:14}) is satisfied).

\section{Holographic energy density and $Q$-function}
\label{sec:IV}

\begin{figure}[H]
 \centering
 \includegraphics[width=6cm,height=6cm]{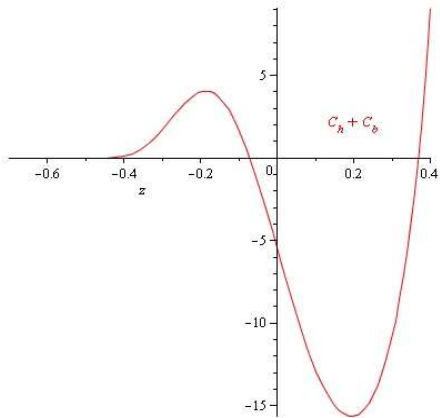}
\caption{We see sign changes in $C_{b}+C_{h}$ for one sign change
in the function $Q$, towards the future the negativity of this sum occurs in 
a narrow range of $z$. Similar behaviours can be observed if we consider two sign changes 
or neither for the function $Q$.}
\label{fig:cero}
\end{figure}

Before introducing an holographic energy density, we start by using the
Ansatz given in Eq. (\ref{eq:1}). In this case, if we use Eq. (\ref{eq:2}) 
we can obtain
\begin{equation}
\frac{Q}{3\eta H^{3}}=2\lambda \left( \frac{q-\frac{1}{2}}{\omega }\right) >0,
\label{eq:15}
\end{equation}
and then, according to Eq. (\ref{eq:8}) we have $\omega _{eff}=\omega +\lambda $. In this
case the bulk temperature becomes
\begin{equation}
T_{b}\left( z\right) = \mathcal{C}_{0}\left( 1+z\right) ^{3\lambda }\exp \left( 3\int 
\frac{dz}{1+z}\omega \right),
\label{eq:16} 
\end{equation}
where $\mathcal{C}_{0}$ is a constant. Now, for the energy density we use the holographic model given by \cite{Granda11}
\begin{equation}
\rho \left( z\right) =3\left[ \alpha -\beta \left( 1+q\right) \right]
H^{2}\left( z\right).
\label{eq:17} 
\end{equation}
Replacing this last expression in Eq. (\ref{eq:3}), after a straightforward calculation we obtain
\begin{widetext}
\begin{equation}
\frac{Q}{3\eta H^{3}}\left( z\right) =-\left( 2\left[ 1-\frac{\alpha}{\eta} +\left(
\frac{\beta}{\eta} \right) \left( 1+q\right) \right] \left( q-\frac{1}{2}\right) +\left(
\frac{\beta}{\eta} \right) \left( 1+z\right) \frac{dq}{dz}\right).  
\label{eq:18}
\end{equation}
\end{widetext}
In Ref. \cite{Lepe2} was shown that $Q$, given in Eq. (\ref{eq:18}), experiences sign
changes through the evolution. For instance, by using the aforementioned $q\left( z\right)$-parametrization 
we can see explicitly those sign changes.
Also, from Eqs. (\ref{eq:2}) and (\ref{eq:17}) we can obtain the relation
\begin{eqnarray}
\frac{q-\frac{1}{2}}{\omega }&=&\frac{3}{2}\left[ \frac{\alpha}{\eta} -\left( \frac{\beta}{\eta}
\right) \left( 1+q\right) \right] \rightarrow \omega = \nonumber \\
&=&\frac{2}{3}\left( 
\frac{q-\frac{1}{2}}{\frac{\alpha}{\eta} -\left(\frac{\beta}{\eta} \right) \left( 1+q\right) }
\right).
\label{eq:19} 
\end{eqnarray}
Finally, the horizon temperature can be obtained by replacing the Eq. (\ref{eq:17}) in Eq. (\ref{eq:2}), so that
we can obtain (by recalling that $T_{h}=H/2\pi $)
\begin{equation}
d\ln \left[\frac{T_{h}\left( z\right)}{T_{h}\left( 0\right)} \right]=\left( \frac{
1+\left(\frac{\alpha}{\eta}\right) \omega \left( z\right) }{1+\left(\frac{3\beta}{2\eta} 
\right) \omega \left( z\right) }\right)d\ln \left( 1+z\right) ^{3/2},
\label{eq:20} 
\end{equation}
and if we consider Eq. (\ref{eq:19}) alongside the $q$-parametrization given before, we
obtain
\begin{equation}
\omega \left( z\right) =\frac{2q_{1}}{3}\frac{z-\bar{z}}{\left(\frac{\alpha}{\eta}
-\frac{3\beta}{2\eta} \right) \left( 1+z\right) ^{2}-\left(\frac{\beta}{\eta} \right)
q_{1}\left( z-\bar{z}\right) },
\label{eq:21} 
\end{equation}%
and then, the solution for $T_{h}$ is given in Eq. (\ref{eq:12}). We end this section 
with a brief discussion about thermal equilibrium between the bulk and the boundary. 
When we think about thermal equilibrium, we do it in the framework of classical thermodynamics, 
where the thermal equilibrium is reached when both temperatures are equal and keep equal throughout the 
evolution. However, in this case the thermal equilibrium appears difficult to achieve. A thermodynamical 
criterion to visualize if the thermal equilibrium is kept once is reached, is verifying that sum of both
heat capacities (bulk and boundary) keeps negative. Given that the heat capacity of the horizon (boundary) 
is negative always $C_{h} = T_{h}\left(\partial S_{h}/\partial T_{h} \right) = -4/T^{2}_{h}$, in Fig. (\ref{fig:cero}) we
observe sign changes in the sum of the heat capacities of the bulk and boundary, $C_{b}$ and $C_{h}$ respectively, this is
a signal of non-thermal equilibrium. The negativity of the sum extends somewhat towards the near future and then experiences
a sign change as well as in the recent past, here we have used the function $Q$ with one change of sign through the evolution.
A similar behaviour can be observed when we consider a function $Q$ with two sign changes or neither. So, according to
Fig. (\ref{fig:cero}), the thermal equilibrium (negativity of $C_{h} + C_{b}$) can be seen as a transient stage, at least, in the
framework of the present discussion. Nothing else can be said. It is difficult conceiving thermal equilibrium also, if, for instance 
we are thinking about the (quantum) concept of entanglement: bulk and boundary (seen as two systems in interaction) should reach the
thermal equilibrium and, sooner or later, maintain it together with the growth of the entanglement entropy, but, 
if we look the Ref. \cite{new} we can find that the aforementioned scheme has not been rigorously proven (despite the list of 
evidences given there). By using the Eqs. (\ref{eq:6}), (\ref{eq:7}) and (\ref{eq:8}), it is straightforward to obtain the heat capacity of the bulk, $C_{b} = T_{b}
(\partial S_{b}/\partial T_{b}) = -(6\pi T_{b}T_{h}\omega_{eff})^{-1}f(z)$, where $f(z)$ is given by the r.h.s. of Eq. (\ref{eq:6}), as
shown in Fig. (\ref{fig:cero}) this heat capacity has not a definite sign through the cosmic evolution.

\section{The second law and phase transitions}

\begin{figure}[H]
 \centering
 \includegraphics[width=5cm,height=5cm]{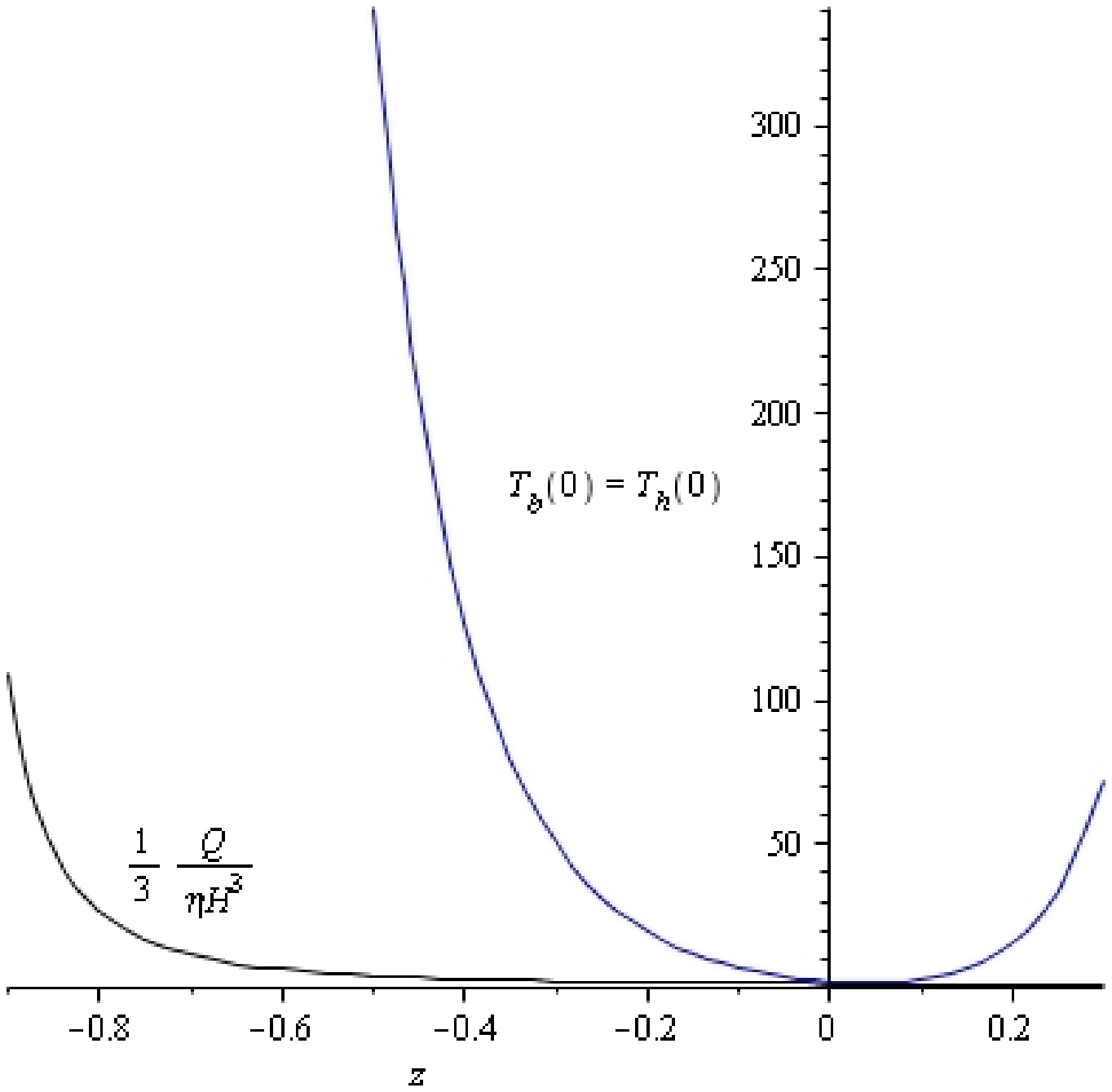}
 \includegraphics[width=5cm,height=5cm]{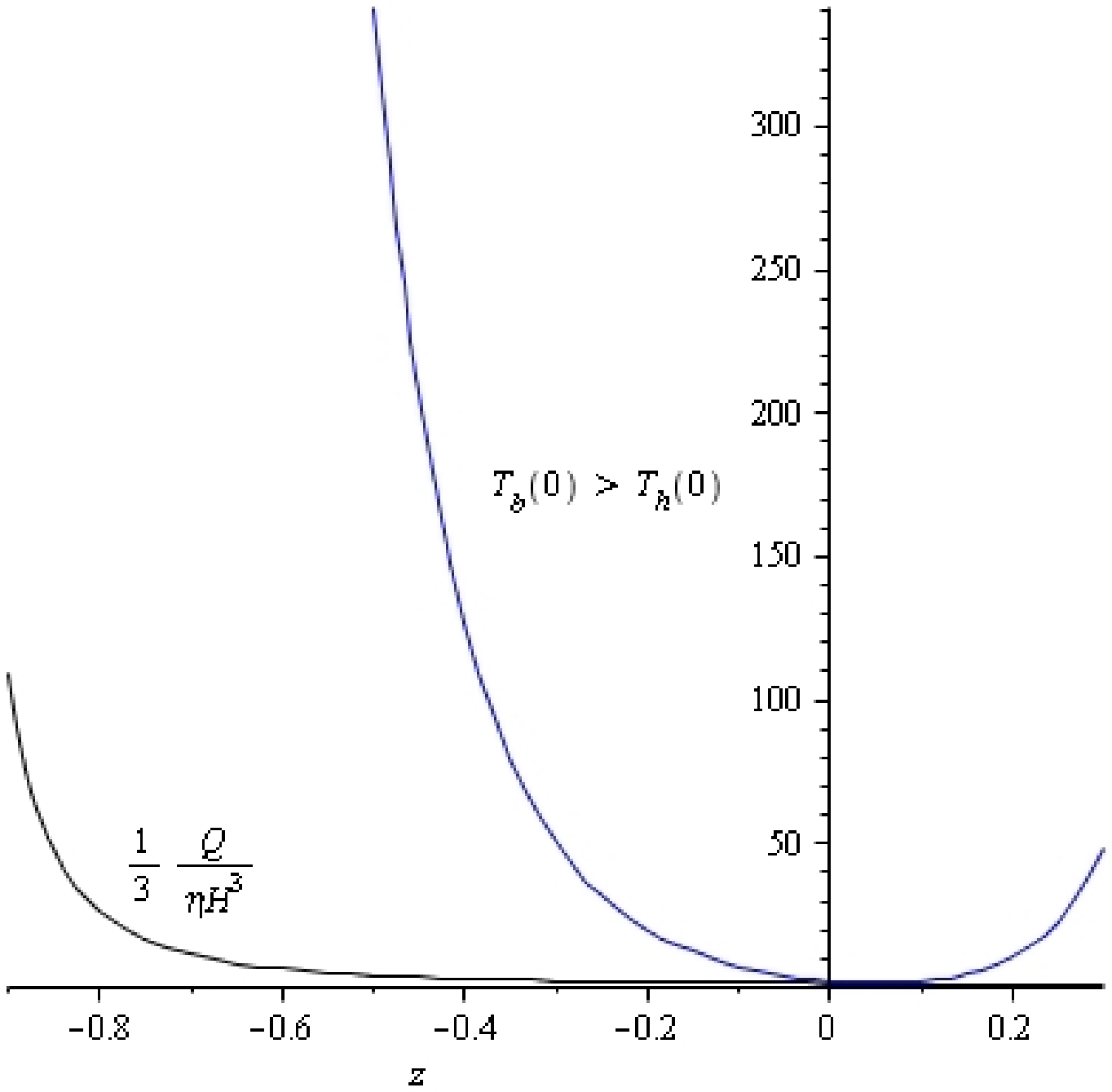}
 \includegraphics[width=5cm,height=5cm]{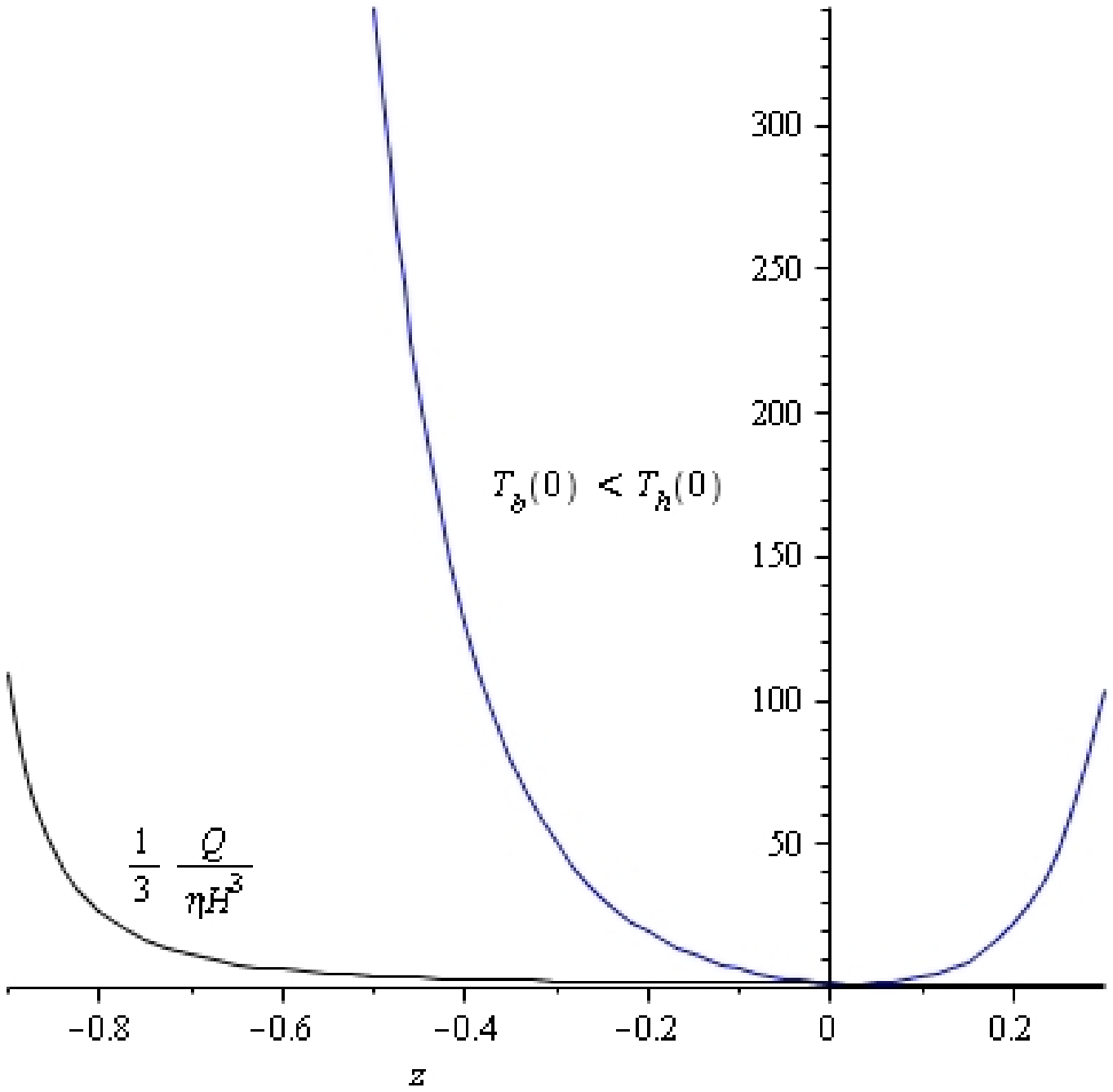}
\caption{According to Eq. (\ref{eq:14}), whose right hand side is represented by the blue line and Eq. (\ref{eq:15}) 
besides the quotient $T_{b}/T_{h}$ given from Eqs. (\ref{eq:16}) and (\ref{eq:12}), we can see 
that the inequality given in Eq. (\ref{eq:14}) is fully satisfied, i.e., the second law is satisfied.}
\label{fig:primera}
\end{figure}

From Fig. (\ref{fig:primera}), we can see that for $Q>0$, the second law it is always
satisfied, independently of the following conditions: $T_{b}\left( 0\right) 
<T_{h}\left( 0\right) $, $T_{b}\left( 0\right) =T_{h}\left( 0\right) $ or $T_{b}\left( 0\right)
>T_{h}\left( 0\right) $. Recent results given by the observational data indicate that $%
Q\neq 0$, but, the question if the function $Q$ is always positive is still open.

\begin{figure}[H]
 \centering
 \includegraphics[width=5cm,height=5cm]{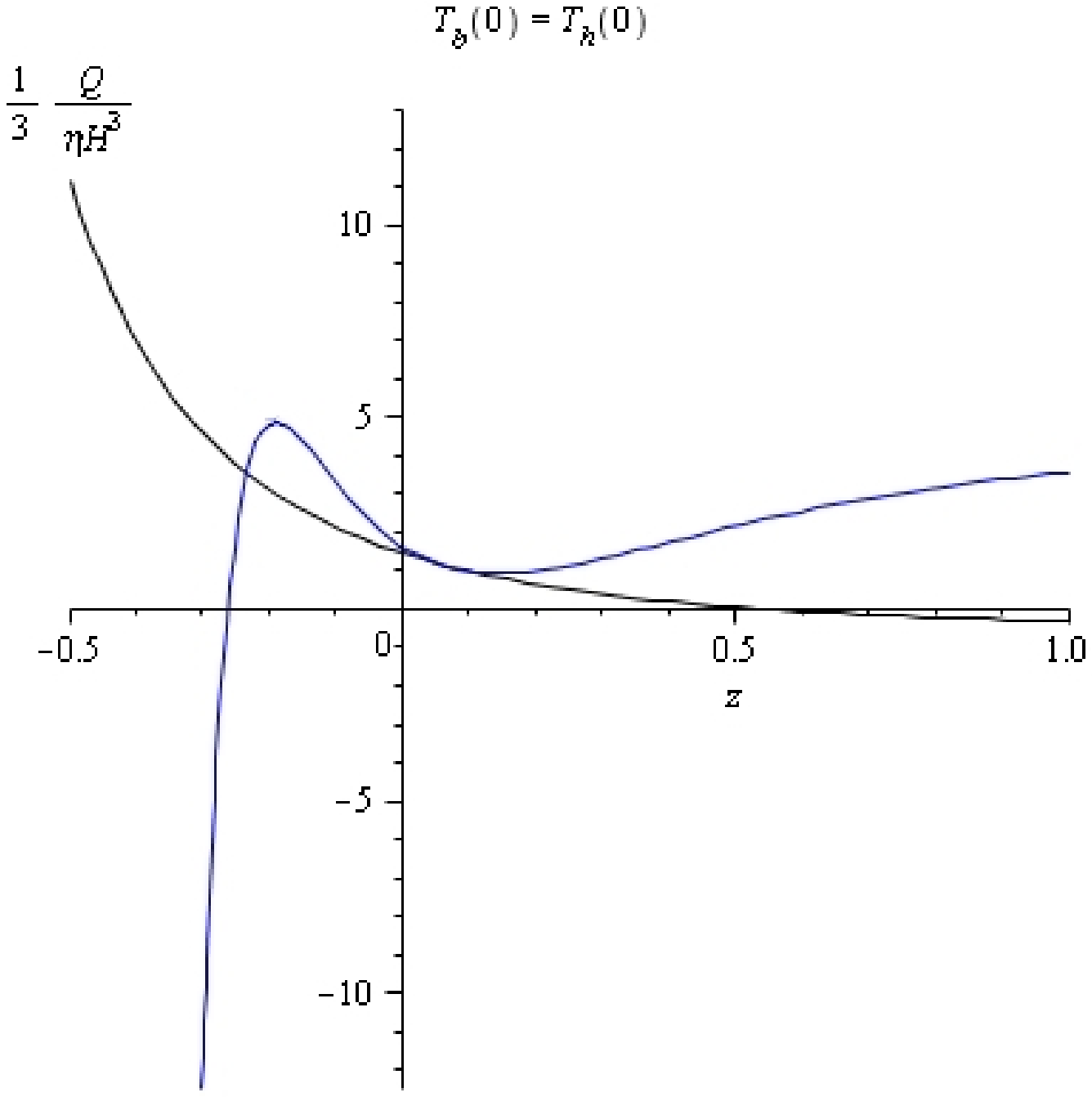}
 \includegraphics[width=5cm,height=5cm]{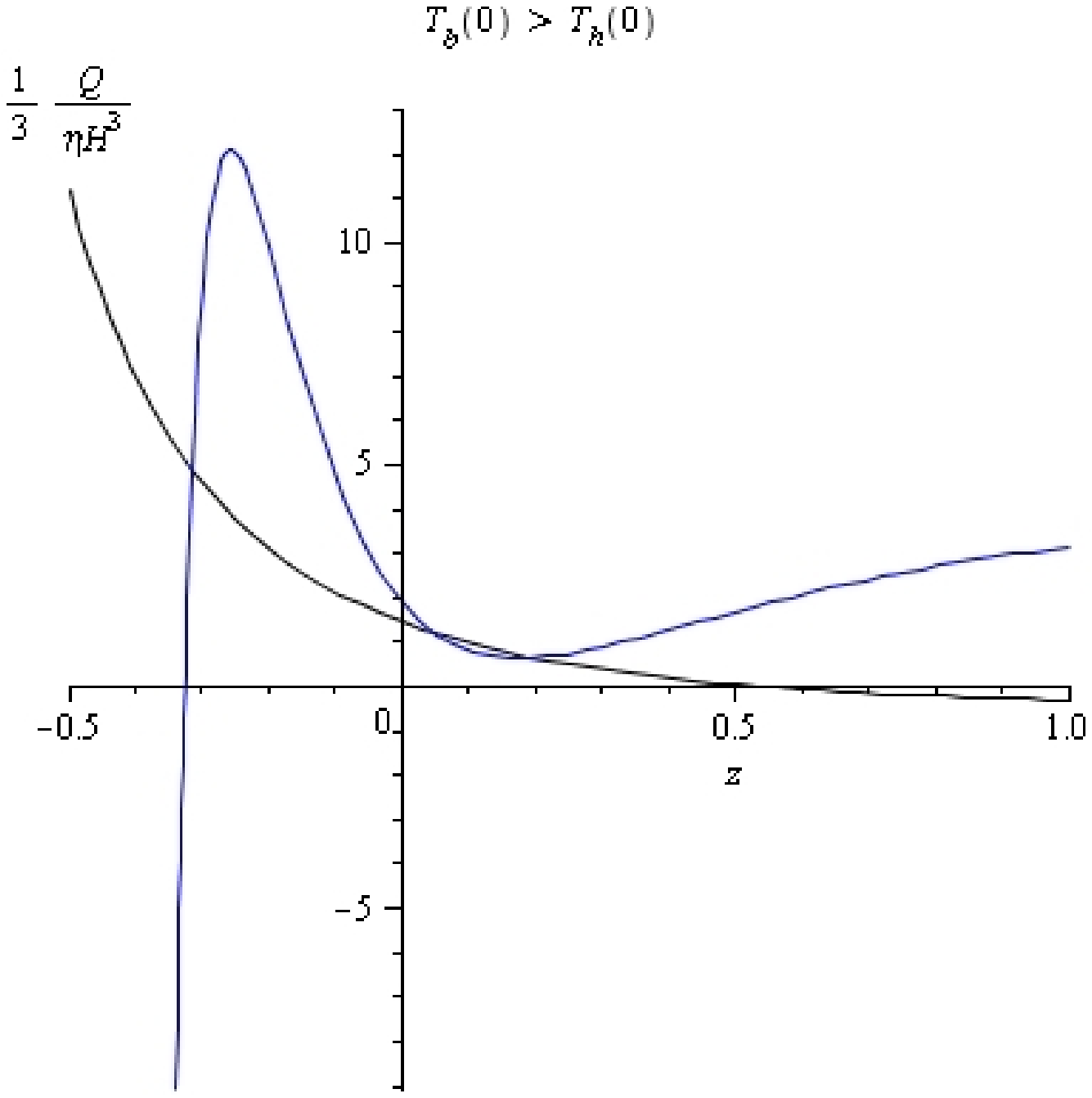}
 \includegraphics[width=5cm,height=5cm]{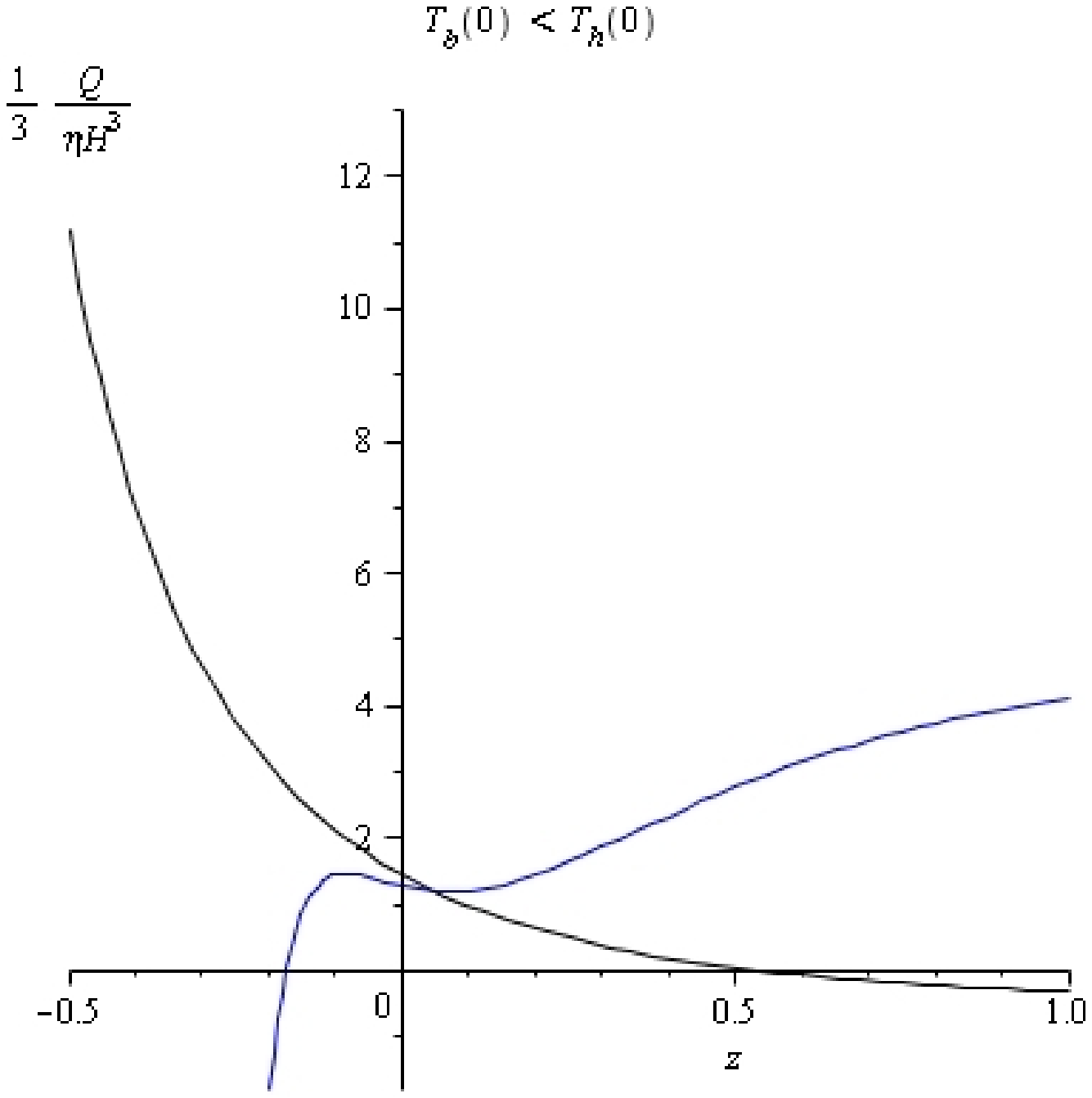}
\caption{According to Eqs. (\ref{eq:14}) and (\ref{eq:18}), with $\beta =0$, and the quotient $%
T_{b}/T_{h}$ given from Eqs. (\ref{eq:9}) and (\ref{eq:12}), we can see regions where the second law is violated and this 
occurs independently of the values of $T_{b}\left( 0\right)$ and $T_{h}\left( 0\right)$. As in Fig. (\ref{fig:primera}) 
we represent the right hand side of Eq. (\ref{eq:14}) by a blue line.}
\label{fig:segunda}
\end{figure}

If we consider one sign change (one zero) in the function $Q$, from Fig. (\ref{fig:segunda}), we can
visualize a strong dependence on the values of $T_{b}\left( 0\right) $ and $%
T_{h}\left( 0\right) $ at time of speaking on the validity of the second
law. For the case when $T_{b}\left( 0\right) <T_{h}\left( 0\right) $ we observe a
violation region from $z_{0}\approx 0.047$ towards the future and a non-violation region
from $z_{0}$ to the past. When $T_{b}\left( 0\right)
=T_{h}\left( 0\right)$, we observe a violation region of the second law at the
future, and finally, for the case when $T_{b}\left( 0\right) >T_{h}\left( 0\right) $
we observe a narrow region in the recent past $ 0.047 \leq z\leq
0.19$ where the second law is violated.\\
Based on some results exposed in Appendix (\ref{sec:app}), for the case shown in Fig. (\ref{fig:tercera}), 
we can observe that second law is satisfied, nevertheless 
the sign (three zeros) of $Q$ changes. What does it mean? Whatever it is, in absence of
observational data for both temperatures, $T_{b}\left( 0\right) $ and $%
T_{h}\left( 0\right) $, we can not say anything conclusive yet.\\
So, apparently, we have problems with the second law only when $Q$ experiences one sign 
change. In other words, everything seems to indicate that the second law is fulfilled if 
the function $Q$ keeps positive. But in the case shown in Fig. (\ref{fig:tercera}), the second law is also 
satisfied although $Q$ experiences three sign changes. This fact redounds what has been said, 
this is, based on this analysis we can not say anything conclusive.

\begin{figure}[H]
 \centering
 \includegraphics[width=6cm,height=5.5cm]{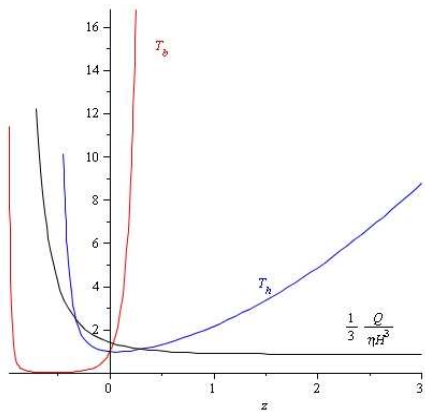}
\caption{According to Eqs. (\ref{eq:15}), (\ref{eq:9}) and (\ref{eq:12}), we can observe a possible
phase transition in the future, the bulk temperature ($T_{b}$) increases even if the
bulk energy decreases (here, $Q>0$ always). The blue line corresponds to the horizon 
temperature ($T_{h}$).}
\label{fig:cuarta}
\end{figure}

\begin{figure}[H]
 \centering
 \includegraphics[width=5cm,height=5cm]{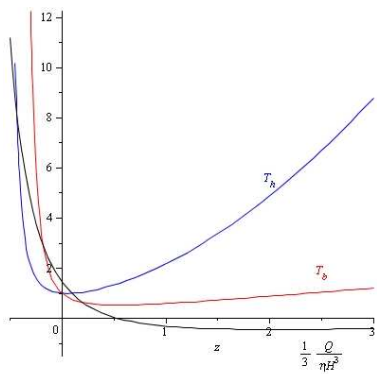}
\caption{According to Eq. (18), $\beta =0$, and Eqs. (\ref{eq:9}) and (\ref{eq:12}). We can see a sign change
in\ $Q$ experienced in the past ($z=\bar{z}$ $=0.54$) and we can see a
possible phase transition for $0\lesssim z\leq $ $\bar{z}$ namely, the bulk
temperature ($T_{b}$) increases and the horizon temperature ($T_{h}$) decreases while $Q>0$.
From $z=0$ to the future the bulk temperature continues growing as well as the horizon temperature.}
\label{fig:quinta}
\end{figure}

\begin{figure}[H]
 \centering
 \includegraphics[width=5cm,height=5cm]{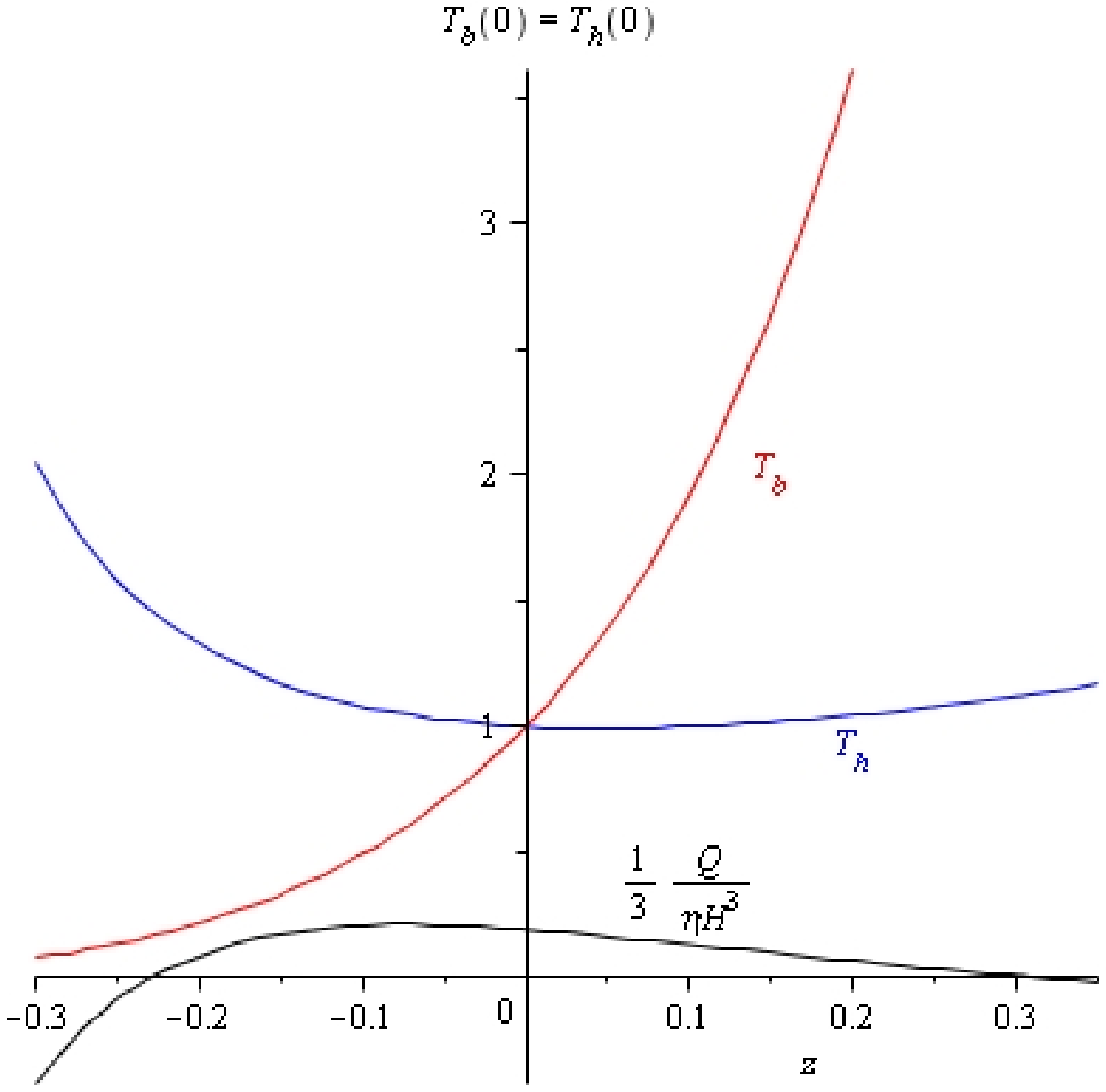}
 \includegraphics[width=5cm,height=5cm]{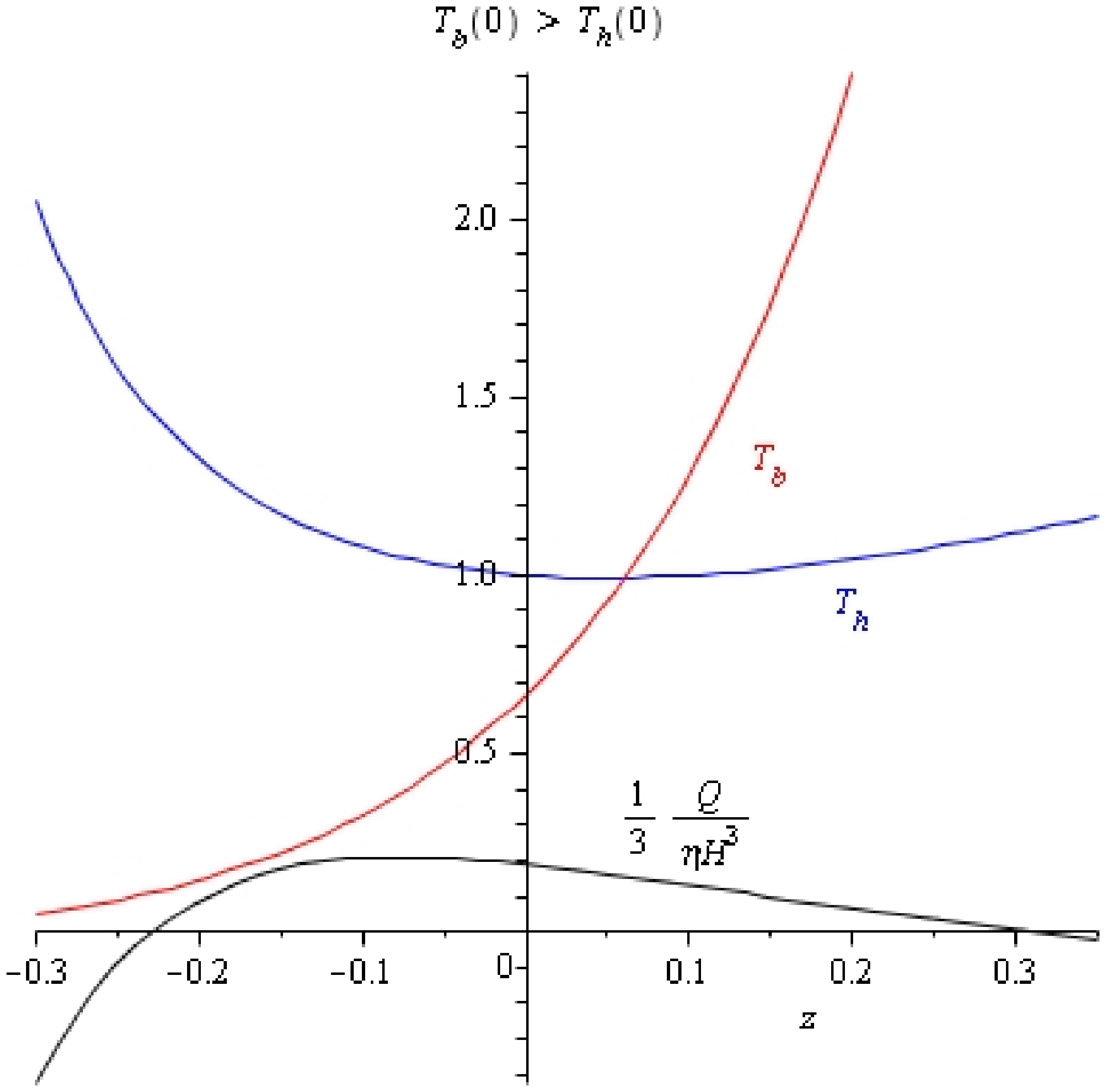}
 \includegraphics[width=5cm,height=5cm]{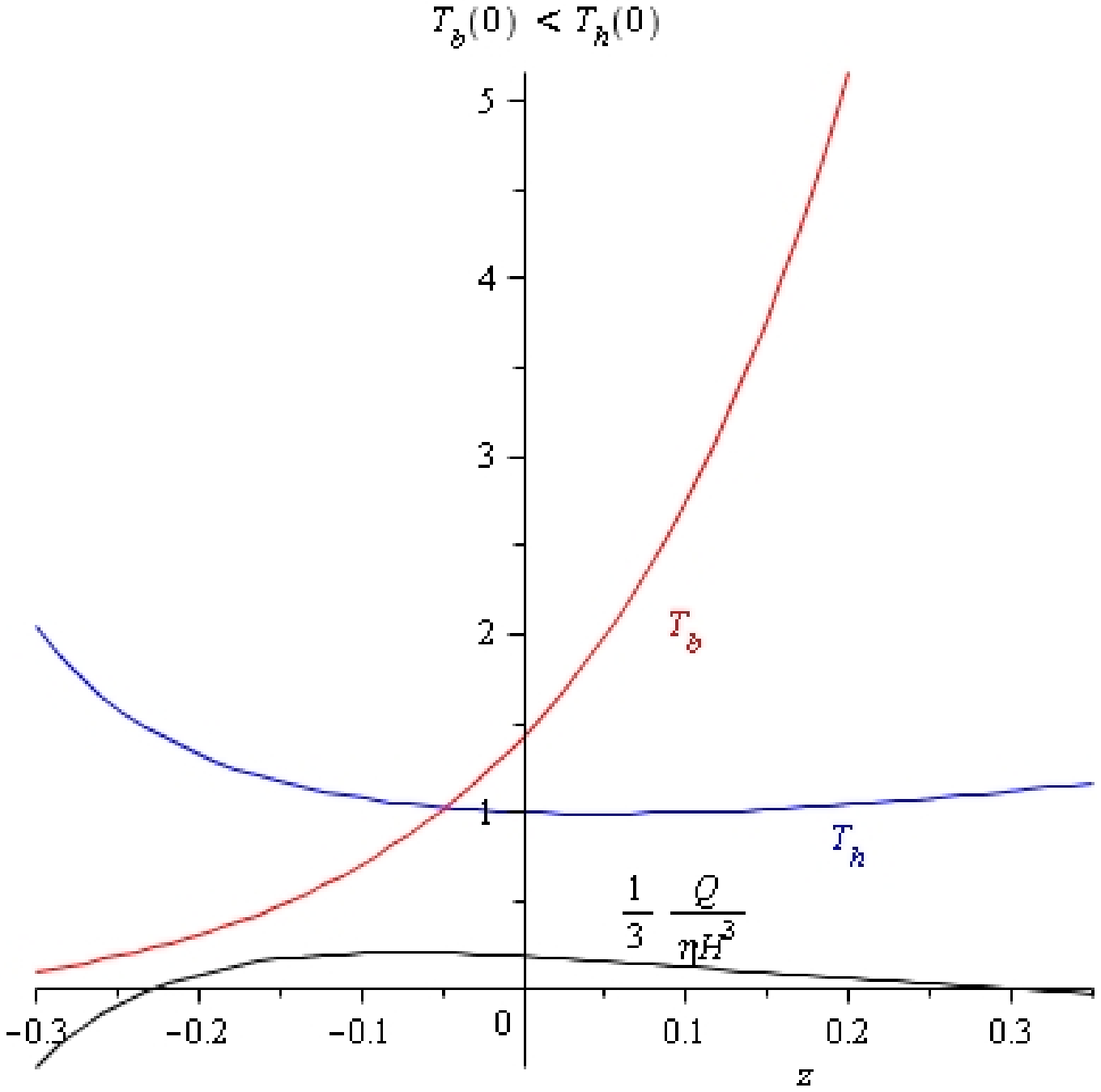}
\caption{According to Eqs. (\ref{eq:18}), (\ref{eq:9}) and (\ref{eq:12}), we have plotted the zone around $%
z=0 $, where we can see two sign changes of $Q$. Here, we do not observe
phase transitions, independently of the values of $T_{b}\left( 0\right) $
and $T_{h}\left( 0\right) $, at least where $Q>0$.}
\label{fig:sexta}
\end{figure}

Finally, according to Figs. (\ref{fig:cuarta}) and (\ref{fig:quinta}), we can visualize phase transitions,
independently of the behaviour of $Q$ (with or without sign changes). This fact is
itself interesting to study, at least in HL cosmology. Finally, in Fig. (\ref{fig:sexta}) at
least where $Q>0$, we do not observe phase transitions.

\subsection{Plots}

All the involved parameters are taken from Ref. \cite{Lepe2}: $\alpha
/\eta =0.6$, $\beta /\eta =0.1$, $q_{1}=3.36$ and $\bar{z}=0.54$, and in
(\ref{eq:16}) we have used $0.6<\lambda <1$. In all cases, we have used the $%
\omega $-parametrization given in (\ref{eq:21}). Additionally, we have added an
Appendix where we show explicitly the involved integrals expressed all in
terms of elemental functions.
\onecolumngrid

\begin{figure}[H]
 \centering
 \includegraphics[width=12cm,height=5cm]{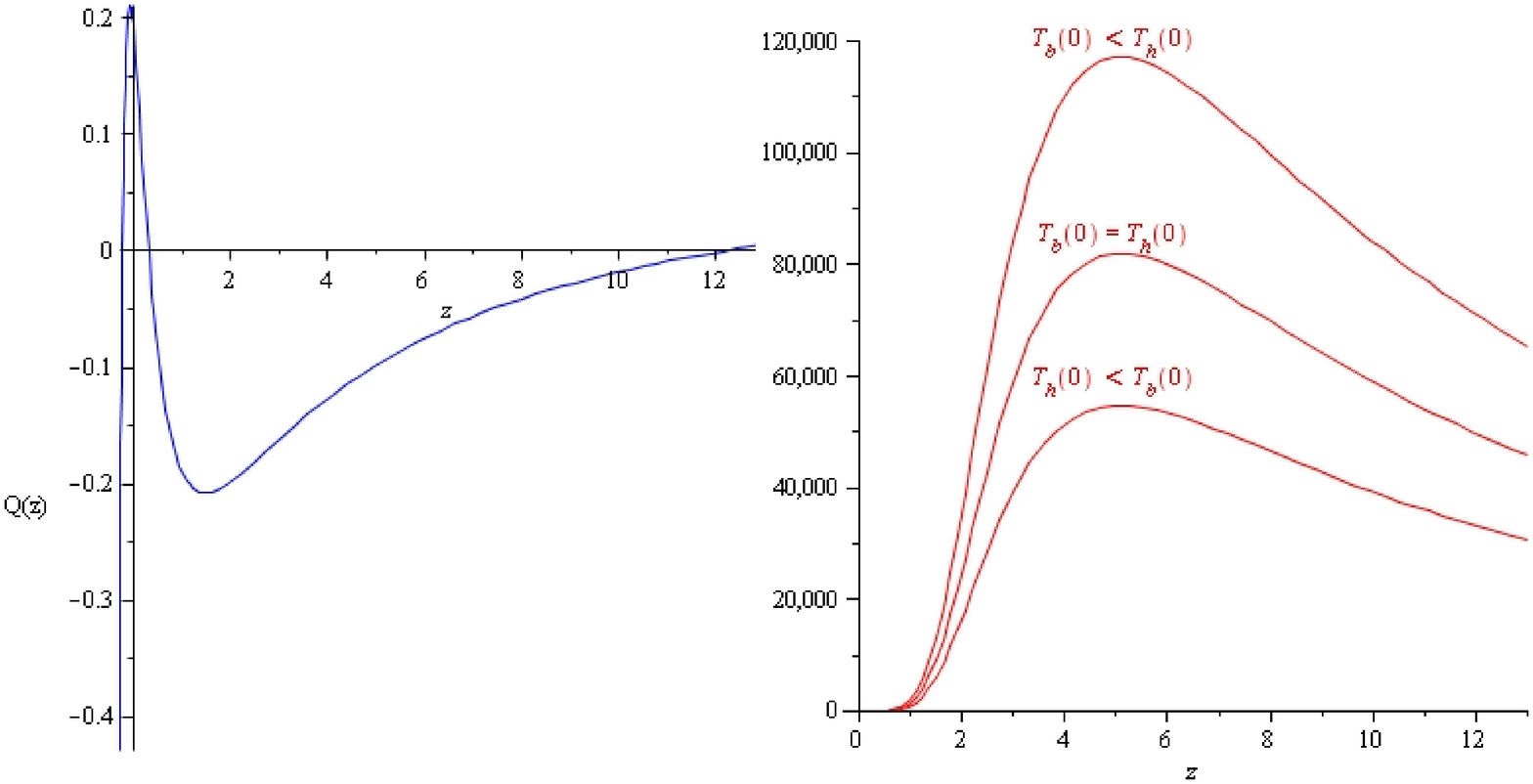}
\caption{According to Eq. (\ref{eq:14}) and the expression for $Q$ given in (\ref{eq:18}), the quotient 
$T_{b}/T_{h}$ given in Eqs. (\ref{eq:9}) and (\ref{eq:12}). In this case, the second law is also satisfied, the right hand side
of Eq. (\ref{eq:14}) is represented in the right hand side plot. The behaviour of $Q$ and its three
signs changes: $Q\left( -0.2<z<0.38\right) >0$, $Q\left( 0.38<z<12.2\right)
<0 $ and $Q\left( z>12.2\right) >0$. We note that $Q\left( z\rightarrow
\infty \right) \rightarrow 0$ and $Q\left( z\rightarrow -1\right) $
diverges. A maximum of $Q\left( z\right) $ is localized around $z=0$.}
\label{fig:tercera}
\end{figure}

\twocolumngrid

\section{Final remarks}

We have discussed the second law in HL cosmology by using a $Q$-interaction
function, energy transference between the bulk and the boundary of the
spacetime besides an holographic model for the energy density in the bulk. By
using a $Q$-function which does not experience sign changes through the
cosmic evolution, we have verified the second law, but, after using a $Q$%
-function which experiences one sign change, we observe problems with the
second law, this is, the second law is verified except in a narrow region in
the recent past $0.047\leq z\leq 0.19$ when we consider $T_{b}\left(
0\right) >T_{h}\left( 0\right) $, the second law is verified except in the
future when we consider $T_{b}\left( 0\right) =T_{h}\left( 0\right) $ and,
if we consider $T_{b}\left( 0\right) <T_{h}\left( 0\right) $, we observe a
violation of the second law at the region $z<0.047$. But, when we consider a $%
Q $-function which experiences three sign changes, the second law is
satisfied independently of the values of $T_{b}\left( 0\right) $ and $%
T_{h}\left( 0\right) $.

On the other hand, if we have $Q>0$ always, we can see a phase transition at the
future, namely, the bulk temperature increases even if the bulk is losing
energy. When $Q$ experiences one sign change (in the past), we also observe 
phase transitions towards the future. And, when $Q$ experiences three sign
changes, we do no not observe phase transitions.

So, if the second law must be respected during cosmic evolution even in
presence of interaction, then, according to what is shown here, $Q$ should
not change sign (or whether, if we take into account the case with three
sign changes of $Q$?). Nevertheless, our work and conclusions are fully
dependent of the observational data. $Q\neq 0$ is a fact well established by
the observation, but the challenge remains in the possibility of confirming sign 
changes of $Q$ in future observations. If this is so, we should expect very 
interesting consequences for the late cosmology, particularly, if the
second law is or not is verified in our universe.

According to Figs. (\ref{fig:primera})-(\ref{fig:sexta}) we have attempted to give a 
message about the validity of the second law in HL cosmology. To be clear, it is only 
a message. Perhaps, we are {\it setting sail} into a new ocean.

Finally, we have seen that the sum of the heat capacities does not maintain the same 
sign (negative) through the cosmic evolution. This can be interpreted 
as a clear signal of non-equilibrium between bulk and boundary. In this case the 
thermal equilibrium may be a transient stage, we do not know yet. So, in HL cosmology 
under an holographic scheme for the energy density and by considering the bulk-boundary 
interaction, we do not visualize thermal equilibrium.

\section*{Acknowledgments}
This work was supported by Pontificia Universidad Cat\'olica de Valpara\'\i so, 
Proyecto de Postdoctorado DI 2015 (M.C.), PUCV-VRIEA Grant No.
037.448/2015, Pontificia Universidad Cat\'olica de Valpara\'\i so (S.L.)
and DIUFRO Grant No. DI14-0007 of Direcci\'on de Investigaci\'on y
Desarrollo, Universidad de La Frontera (F.P.). The authors acknowledge to Diego 
Pav\'on for his enlightening discussions. M.C. also thank the Instituto
de F\'\i sica de la Pontificia Universidad Cat\'olica de Valpara\'\i so for its 
hospitality during the preparation of this manuscript.

\appendix
\section{Some results for $Q$-function}
\label{sec:app}
According to Eq. (\ref{eq:16}) besides of the given $q$ and $\omega $%
-parametrizations, the bulk temperature is

\begin{eqnarray}
\frac{T_{b}\left( z\right)}{T_{b}\left( 0\right)} &=&\left( 1+z\right) ^{3\lambda
}\exp \left( \frac{2q_{1}}{A}\left( \left[ I_{1}\left( z\right) -I_{1}\left(
0\right) \right] - \right. \right. \nonumber \\
&-& \left. \left. \left( 1+\bar{z}\right) \left[ I_{2}\left( z\right)
-I_{2}\left( 0\right) \right] \right) \right),
\end{eqnarray}%
where
\begin{equation}
I_{1}\left( z\right) =\frac{2}{\sqrt{\Delta }}\arctan \left( \frac{2\left(
1+z\right) -B/A}{\sqrt{\Delta }}\right),
\end{equation}%
and
\begin{equation}
I_{2}\left( z\right) =\frac{1}{C}\left[ \left( \frac{B}{2A}\right) I_{1}\left(
z\right) +\ln \left( \frac{1+z}{\sqrt{R\left( z\right) }}\right) \right], 
\end{equation}%
being $0.6<\lambda <1$, $A=\alpha /\eta -3\beta /2\eta $, $B=\left( \beta
/\eta \right) q_{1}$, $C=\left( 1+\bar{z}\right) \left( B/A\right) $, $%
\Delta =4C-\left( B/A\right) ^{2}$ and $R\left( z\right) =\left( 1+z\right)
^{2}-\left( B/A\right) \left( 1+z\right) +C$. And $\alpha /\eta =0.6$, $%
\beta /\eta =0.1$, $q_{1}=3.36$ and $\bar{z}=0.54$. This case correspond to
use Eq. (\ref{eq:15}) where there is not sign change of $Q$.

$\bullet$ Following Eq. (\ref{eq:18}), if we consider $\beta =0$  we have
\begin{equation}
\frac{Q}{3\eta H^{3}}=-2\left( 1-\frac{\alpha}{\eta} \right) \left( q-\frac{1}{2}\right), 
\end{equation}%
(one sign change in $Q$) and then, according to Eq. (\ref{eq:9}), the bulk temperature is
\begin{align}
& \frac{T_{b}\left( z\right)}{T_{b}\left( 0\right)} =\exp \left[ q_{1}\left( 1-\bar{z}%
\right) \right]\times \nonumber \\
&\times \exp \left[ -2q_{1}\left\lbrace \frac{1}{1+z}-\left( \frac{1+\bar{z%
}}{2}\right) \frac{1}{\left( 1+z\right) ^{2}}\right\rbrace \right].
\end{align}

$\bullet$ Considering Eq. (\ref{eq:18}), and $\alpha \neq 0$ and $\beta \neq 0$, and
according to Eq. (\ref{eq:9}) we have (with $x=1+z$)
\begin{widetext}
\begin{eqnarray}
\frac{T_{b}\left( x\right)}{T_{b}\left( 1\right)} &=&
\exp \left[ \frac{2q_{1}}{A}\left\{ \left( \frac{\alpha}{\eta} -\frac{\beta}{\eta}
\right) \left( I_{1}\left( x\right) -I_{1}\left( 1\right) \right) -\left[
\left( \frac{\alpha}{\eta} -\frac{\beta}{2\eta} \right) \bar{x}+\left( \frac{\beta}{\eta} \right)
q_{1}\right] \left( I_{2}\left( x\right) -I_{2}(1)\right) \right\} \right]
\times  \nonumber \\
&\times & \exp \left[ \frac{2q_{1}^{2}}{A}\left(\frac{\beta}{\eta} \right) \left\lbrace 2%
\bar{x}\left( I_{3}\left( x\right) -I_{3}\left( 1\right) \right) -\bar{x}%
^{2}\left( I_{4}\left( x\right) -I_{4}\left( 1\right) \right) \right\rbrace
\right],
\end{eqnarray}
\end{widetext}
where
\begin{widetext}
\begin{equation}
I_{3}\left( x\right) =-\left( \frac{1}{1+C-\frac{B}{A}}\right) \frac{1}{x}+ 
\frac{\left(\frac{B}{A}-2\right) }{\left( 1+C-\frac{B}{A}\right) ^{2}}\ln \left( \frac{x%
}{\sqrt{AR\left( 1+x\right) }}\right) + 2\left( \frac{(1-C-\left( \frac{B}{A}\right) +\left(\frac{B}{A}\right) ^{2}/2}{\left(
1+C-\frac{B}{A}\right) ^{2}\sqrt{\Delta }}\right) \arctan \left( \frac{2\left(
1+x\right) -\frac{B}{A}}{\sqrt{\Delta }}\right),
\end{equation}
\end{widetext}
and
\begin{widetext}
\begin{eqnarray}
I_{4}\left( x\right)  &=& A\left( \frac{2-\frac{B}{A}}{\left( 1+C-\frac{B}{A}\right) ^{2}}%
\right) \frac{1}{x}-\frac{A}{2}\left( \frac{1}{1+C-\frac{B}{A}}\right) \frac{1}{x^{2}%
}+ 3\left( \frac{1-C/3-\frac{B}{A}+\left( B/A\right) ^{2}/3}{\left( 1+C-\frac{B}{A}\right)
^{3}}\right) \ln \left( \frac{x}{\sqrt{AR\left( 1+x\right) }}\right) + 
\nonumber \\
&+& 3\left( \frac{\left( 1-C\right) \left( \frac{B}{A}\right) -\left( \frac{B}{A}\right)
^{2}+\left( \frac{B}{A}\right) ^{3}/3+2\left( C-1/3\right) }{\left( 1+C-\frac{B}{A}\right)
^{3}\sqrt{\Delta }}\right)\arctan \left( \frac{2\left( 1+x\right) -\frac{B}{A}}{%
\sqrt{\Delta }}\right),
\end{eqnarray}
\end{widetext}
where we have defined 
\begin{equation*}
R\left( 1+x\right) =\left( 1+x\right) ^{2}-\left( \frac{B}{A}\right)
\left( 1+x\right) +C.
\end{equation*}

\end{document}